\documentclass[a4paper,11pt]{article}
\usepackage{jinstpub}
\usepackage{lineno}
\usepackage{makecell}
\usepackage{subcaption}

\title{\boldmath Evaluation of CZT Detector Performance in Alpha and Gamma Spectrometry}

\author[a,b,c,*]{N.~Kramarenko,\note[*]{Corresponding author.}}
\author[a,b,c]{M.~Väänänen,}
\author[a,c]{M.~Kalliokoski,}
\author[a,b,c]{M.~Bezak,}
\author[a,c]{R.~Turpeinen}
\author[d]{and A.~Winkler}
\affiliation[a]{Helsinki Institute of Physics,\\Gustaf Hällströmin katu 2, FI-00014 University of Helsinki, Finland}
\affiliation[b]{Lappeenranta-Lahti University of Technology LUT,\\Yliopistonkatu 34, FI-53850 Lappeenranta, Finland}
\affiliation[c]{University of Helsinki, Department of Physics,\\Gustaf Hällströmin katu 2, FI-00014 University of Helsinki, Finland}
\affiliation[d]{Detection Technology Plc,\\A Grid, Otakaari 5a, FI-02150 Espoo, Finland}

\emailAdd{nikita.kramarenko@helsinki.fi}

\abstract{Presented is a work on the performance studies of Cadmium Zinc Telluride (CZT) planar detector structures. Current–voltage (IV), capacitance–voltage (CV), and gamma- and alpha-spectroscopy measurements were carried out to provide the essential baseline required for forthcoming defect configuration impact studies on the detector spectroscopy performance. For each tested sensor, spectroscopic responses were recorded with several bias voltages applied. The described suite of measurements provides the parameters needed to evaluate, bulk resistivity, signal efficiency, and energy resolution for characteristic peaks at different energies. Readout configurations and data processing are discussed in related sections.}


\arxivnumber{2509.26437v2}

\begin{document}
\maketitle
\flushbottom

\section{Introduction}
\label{sec:Intro}

Cadmium zinc telluride is a room-temperature semiconductor widely used in medical physics, nuclear security, and X-ray, gamma-ray and, less commonly, alpha spectroscopy~\cite{a}. Despite established applications, the processing and characterization of CZT detectors remain complex, with performance being affected by the material surface and bulk defects. In this work a comprehensive characterization of CZT planar structures is presented to establish a performance baseline. The suite of measurements employed provides a solid basis for future studies that aim to better understand the link between the defect composition and the spectroscopic performance, ultimately providing feedback to the crystal growth process.

\section{Setup and Electronics}
\label{sec:Setup}

Five planar CZT sensors grown by the Vertical Gradient Freeze (VGF) method were studied. Each sensor had 100~nm thick gold electrodes at both interfaces, the geometrical parameters of the samples are shown in table \ref{tab:parameters}. Electrical connections were established by attaching bonding wires to the gold electrodes with conductive adhesive and wire-bonding them to the Printed Circuit Board (PCB) pads. The described assembly resulted in a semi-Ohmic contact configuration~\cite{b}. This introduces a parasitic capacitance that is small compared to the bulk capacitance. The PCBs were placed inside a metal enclosure. The top cover included a 3 mm thick lead layer with a centered, 1 mm diameter collimator opening. The PCB hosting the Detector Under Test (DUT) is placed on the 3D printed holder, keeping the detector lined up underneath the collimator. The signal line from the DUT to the first amplification stage was kept as short as possible to reduce induced noise.

\begin{table}[htbp]
\centering
\caption{Sample geometrical parameters.\label{tab:parameters}}
\smallskip
\begin{tabular}{|c|c|c|}
\hline
Sample & \makecell{Thickness (mm)}& \makecell{Electrode area ($\mathrm{mm^2}$)}\\
\hline
1 & 1.33 & 24.40\\
\hline
2 & 1.70 & 23.77\\
\hline
3 & 2.00 & 24.16\\
\hline
4 & 1.60 & 25.00\\
\hline
5 & 1.45 & 18.06\\
\hline
\end{tabular}
\end{table}

\begin{figure}[htbp]
    \centering

    \begin{subfigure}[b]{0.65\textwidth}
        \centering
        \includegraphics[width=\textwidth]{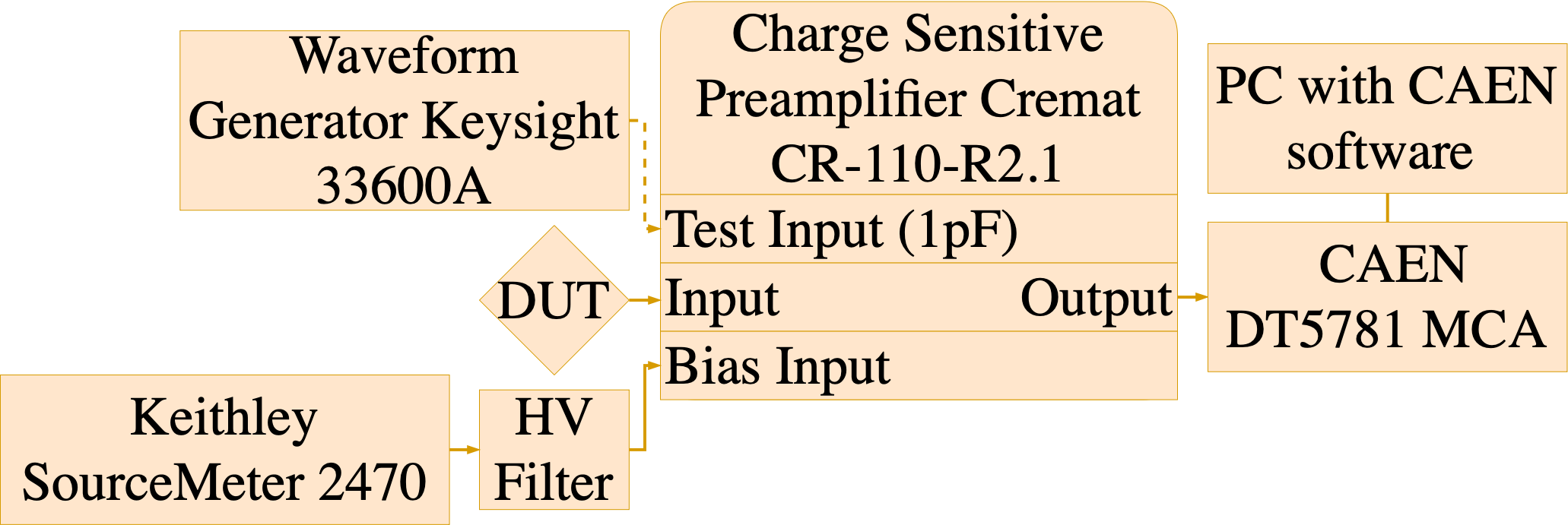}
        \caption{}
        \label{fig:BD1}
    \end{subfigure}

    \begin{subfigure}[b]{0.65\textwidth}
        \centering
        \includegraphics[width=\textwidth]{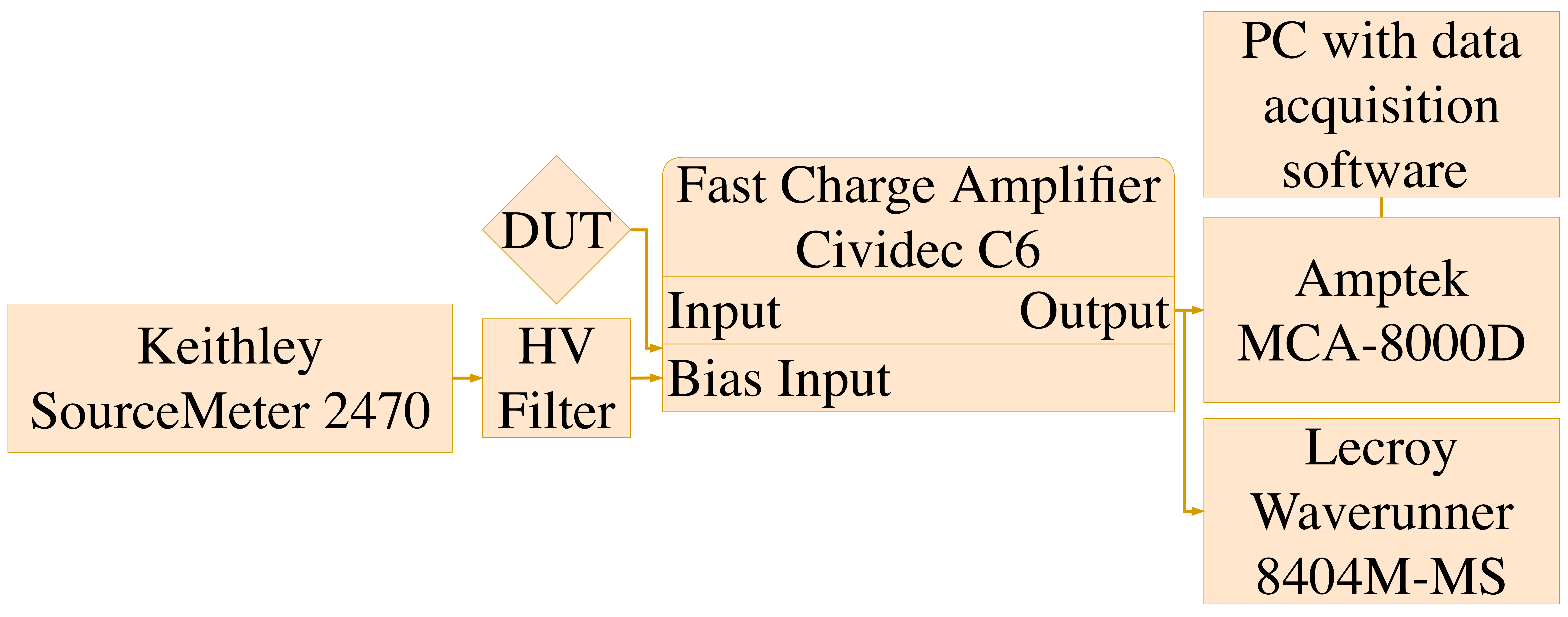}
        \caption{}
        \label{fig:BD2}
    \end{subfigure}
    \caption{Block diagrams illustrating two different data acquisition setups used. (a) Setup for characteristic peak position - voltage dependence and spectra measurements with gamma radiation sources. (b) Setup for the alpha source spectra and waveform measurements.}
    \label{fig:BD_combined}
\end{figure}

The electrical behaviors reflected in the IV and CV characteristics were studied by using Keithley 2410 source meter, Keithley 6487 picoammeter and Agilent E4980A LCR meter. Two irradiation setups were developed to study the response of the samples, as schematically illustrated in figure \ref{fig:BD_combined}.

To test the gamma response of the samples, the enclosure was connected to the readout amplification chain featuring the Cremat CR-150-R5 CSP evaluation board with the Cremat CR-111-R2.1 Charge Sensitive Preamplifier (CSP) module and CAEN DT5781 Multi-Channel Analyzer (MCA) acting as a shaper along its MCA functions. The MCA parameters (rise time, decay time, trigger threshold, trigger hold-off) were tuned based on the preamplifier step-like pulse response. CAEN MCA gain was set to the highest value of 10 and the range was set to 0.3 V peak to peak (Vpp). The MCA response to the pulse generator signals was recorded and shown to be linear over the 0.3 Vpp voltage range.

Spectroscopy and waveform measurements with an alpha source were carried out by using the Cividec C6 fast charge amplifier, which includes a built-in shaper. It was connected to the Amptek MCA-8000D. The internal logic of the MCA is optimized for positive peaks and expects unipolar pulses. Thus, a positive offset is required to detect negative signals. Since the employed measurement setup and bias configuration produced positive pulses at the Amptek MCA-8000D input, no further adjustments were required.

Collimated $^{241}\text{Am}$ with the main photopeak emission at 59.5 keV and activity of $A(t) \approx 397.0 \, \text{kBq}$; $^{133}\text{Ba}$ (356.0 keV, $A(t) \approx 40.9 \, \text{kBq}$); and $^{137}\text{Cs}$ (661.7 keV, $A(t) \approx 98.5 \, \text{kBq}$) were used. The tri-alpha source with an effective activity of $A(t) \approx 10 \, \text{kBq}$ was used. The source contains three isotopes: $^{239}\text{Pu}$ (alpha emission at 5.24 MeV), $^{241}\text{Am}$ (5.49 MeV), and $^{244}\text{Cm}$ (5.81 MeV). The alpha source has the shape of a thin disk and was collimated with a 2 mm round opening in a 1 mm thick lead cover. 

\section{Electrical characteristics}
\label{sec:IVCV}

\begin{figure}[htbp]
\centering
\includegraphics[width=.82\textwidth]{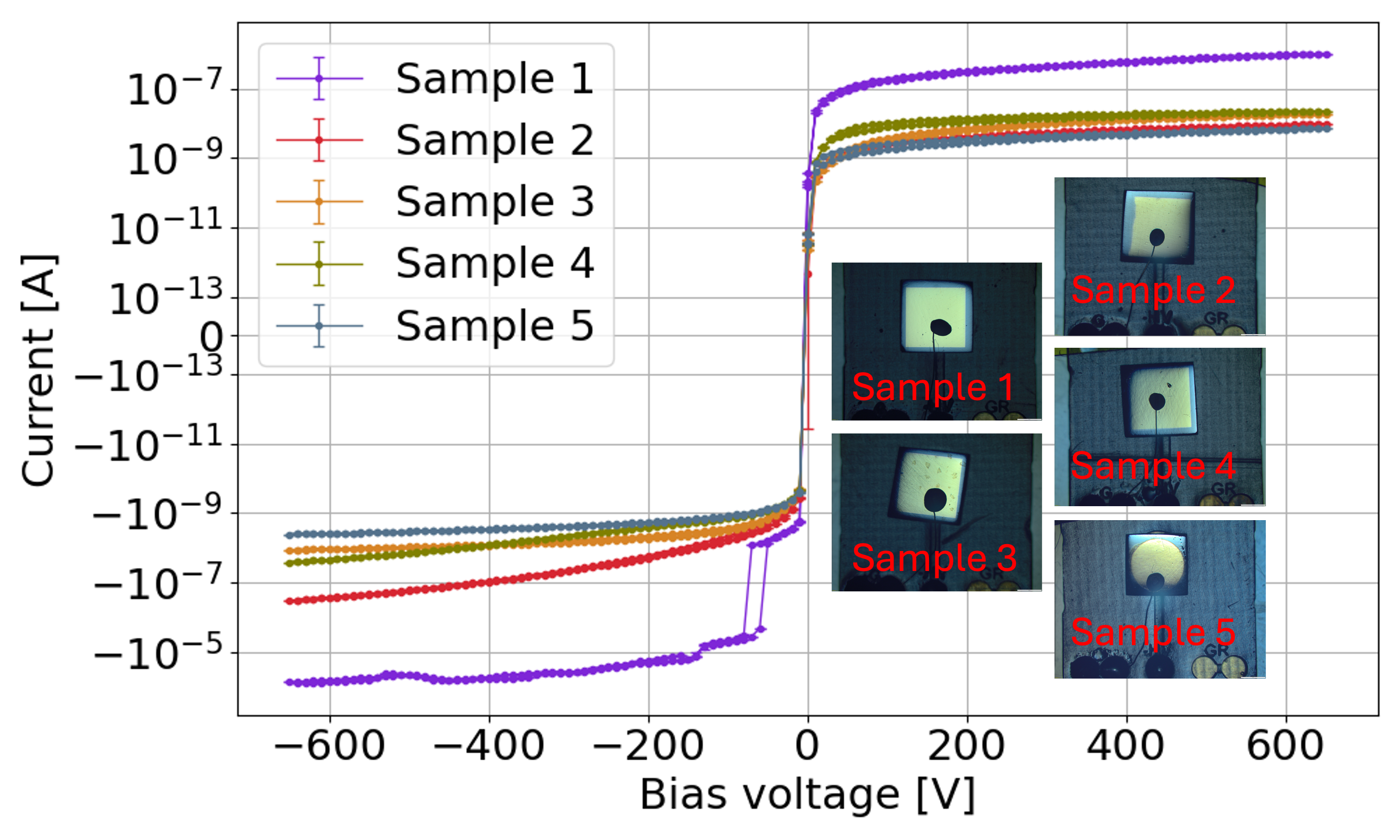}
\caption{IV curves measured for the samples under positive and negative bias voltages in both directions. The curves are plotted using a logarithmic scale for y-axis. Microscope images of the tested samples show the differences in electrode shape configurations.\label{fig:IV}}
\end{figure}

For each sample, the IV and CV characteristics under increasing and decreasing bias voltage (bidirectional scans) in the range from -650 V to 650 V were recorded. The current-voltage characteristics acquired from the front electrodes of the samples are shown in figure \ref{fig:IV}. Bias voltage was applied to the back electrode. For the CV measurements, the bias configuration remained the same.

Sample 1 showed unusual behavior during both IV and CV tests. Hysteresis, as well as non-stability, can be observed, especially when a negative potential is applied to the back electrode. From the IV curves, the resistivity values were calculated using the formula \ref{eq:resistivity} and are shown together with the measured and theoretical capacitance values in table \ref{tab:electrical_behavior}. The studied sensors are assumed to be fully depleted at a 500 V bias, therefore the theoretical capacitance can be approximated as the geometric one of a parallel plate capacitor, see \ref{eq:capacitance}. The measured capacitance values are shown to be in good agreement with the theoretical ones. The round shape of the sample 5 electrode was taken into account in the electrode area and theoretical capacitance calculations.

\begin{equation}
\label{eq:resistivity}
\begin{aligned}
\rho = R_{\text{IV}} \frac{A}{L} \,.
\qquad
\end{aligned}
\end{equation}
\noindent The resistivity $\rho$ is calculated using the bulk resistance $R_{\text{IV}}$, which is determined from the linear region of the measured current-voltage characteristics.

\begin{equation}
\label{eq:capacitance}
\begin{aligned}
C &= \frac{\epsilon_0 \epsilon_{r, \text{CZT}} A}{L} \,.
\qquad
\end{aligned}
\end{equation}
\noindent where $C$ is the theoretical geometric capacitance, $\epsilon_0$ and $\epsilon_{r, \text{CZT}} = 10.9$~\cite{c} are the vacuum and relative permittivity, $A$ is the contact area, and $L$ is the detector thickness.

\begin{table}[htbp]
\centering
\caption{Sample resistivity and capacitance values.\label{tab:electrical_behavior}}
\smallskip
\begin{tabular}{|c|c|c|c|}
\hline
Sample & \makecell{Resistivity ($\Omega \cdot \mathrm{m}$)} & \makecell{Theoretical C (pF)} & \makecell{Measured C at 500V (pF)}\\
\hline
1 & $8.19 \times 10^{6}$ & 1.77 & $1.7377 \pm 0.0001$\\
\hline
2 & $3.85 \times 10^{8}$ & 1.35 & $1.3503 \pm 0.0002$\\
\hline
3 & $3.72 \times 10^{8}$ & 1.17 & $1.2112 \pm 0.0002$\\
\hline
4 & $9.90 \times 10^{7}$ & 1.51 & $1.4675 \pm 0.0002$\\
\hline
5 & $6.31 \times 10^{8}$ & 1.08 & $1.1412 \pm 0.0002$\\
\hline
\end{tabular}
\end{table}

\section{Response to gamma irradiation}
\label{sec:Gamma}

\paragraph{Energy calibration.} Energy calibration was performed by using the characteristic peaks of $^{133}\text{Ba}$\\ (81 keV and 356 keV) and $^{137}\text{Cs}$ (662 keV) gamma emissions, confirming a linear response over this energy range. The Energy Resolution (ER) for emissions above 100 keV deteriorated significantly, due to incomplete energy deposition resulting from the limited stopping power.

The calibration points to convert the MCA channels to energy were obtained by fitting known characteristic peaks of $^{133}\text{Ba}$ and $^{137}\text{Cs}$ with Gaussian functions. Although the peaks obtained with planar CZT sensors exhibit asymmetry, the error introduced by the Gaussian approximation in determining the peak centroids is minimal.

\paragraph{$^{241}\text{Am}$ spectra.}

\begin{figure}[htbp]
\centering
\includegraphics[width=.82\textwidth]{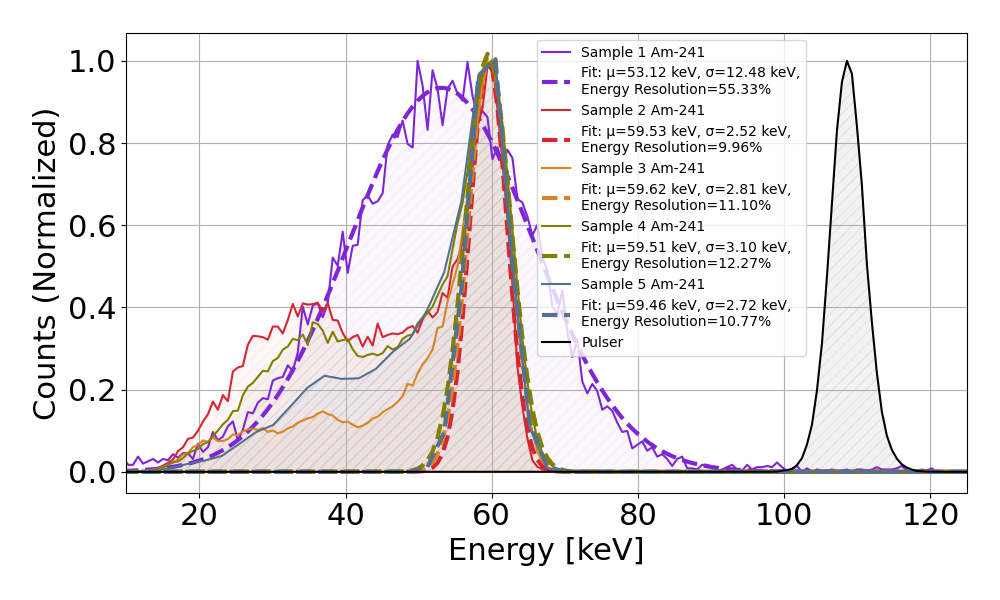}
\caption{$^{241}\text{Am}$ spectra at 650 V bias with peaks fit with a Gaussian function corresponding to the 59.51 keV emission. A 10 mVpp pulse generator signal is shown as a reference. Low-energy channels compromised by electronic noise are excluded from the analysis.\label{fig:Gamma}}
\end{figure}

\begin{figure}[htbp]
\centering
\includegraphics[width=.82\textwidth]{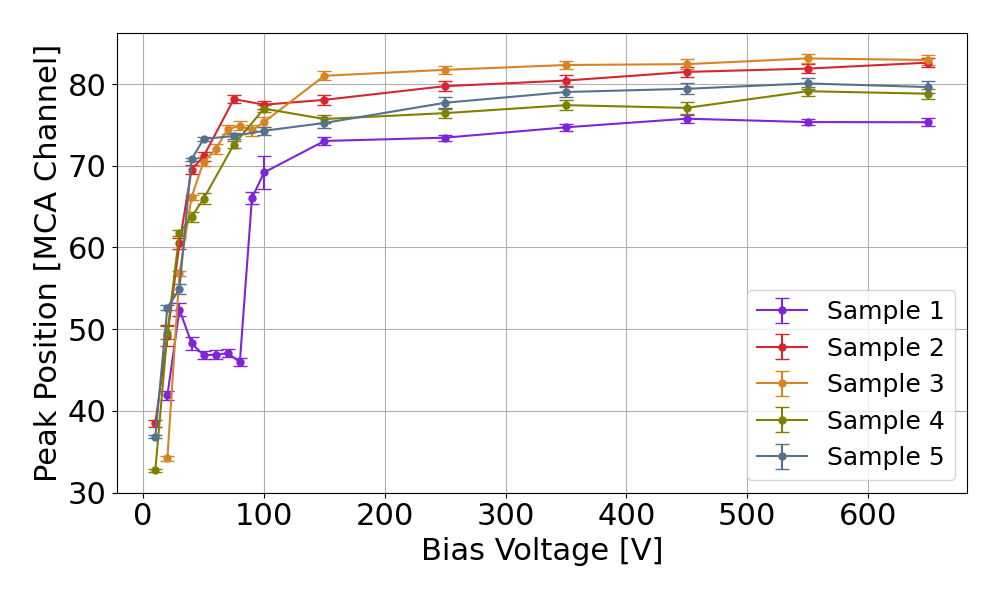}
\caption{Peak position - voltage dependence.\label{fig:PeakPosition}}
\end{figure}

The acquired $^{241}\text{Am}$ spectrum histograms are shown in figure \ref{fig:Gamma}. The voltage dependence of the main $^{241}\text{Am}$ photopeak position was analyzed for the tested structures by calculating the mean and standard deviation of the Gaussian function fits. The resulting curves are shown in figure \ref{fig:PeakPosition}.

As expected, reducing the bias voltage leads to the shift of the photopeak peak centroid toward lower energy channels on the MCA. The response to the 59.51 keV characteristic $^{241}\text{Am}$ gamma emission for sample 1 deviates from the other samples, making it challenging to measure the peak position - voltage dependence at low bias voltage values. Due to the pronounced spectral broadening observed in sample 1, this limitation becomes evident at higher voltages compared to the other samples. Electronic noise introduced to the lower MCA channels for bias voltages below 90 V (for sample 1) alters the emission response histogram shape. One of the methods employed to counteract this adverse effect is to change the Least Significant Bit (LSB) value and then to adjust the filter settings in CAEN MCA such as decay and rise pulse times. Lowering the LSB value leads to an increased sensitivity for low-amplitude signals, including electronic noise, making it difficult to find the balance between the acceptable noise level and the processable detector response.

\section{Response to alpha irradiation}
\label{sec:Alpha}

\begin{figure}[htbp]
\centering
\includegraphics[width=.82\textwidth]{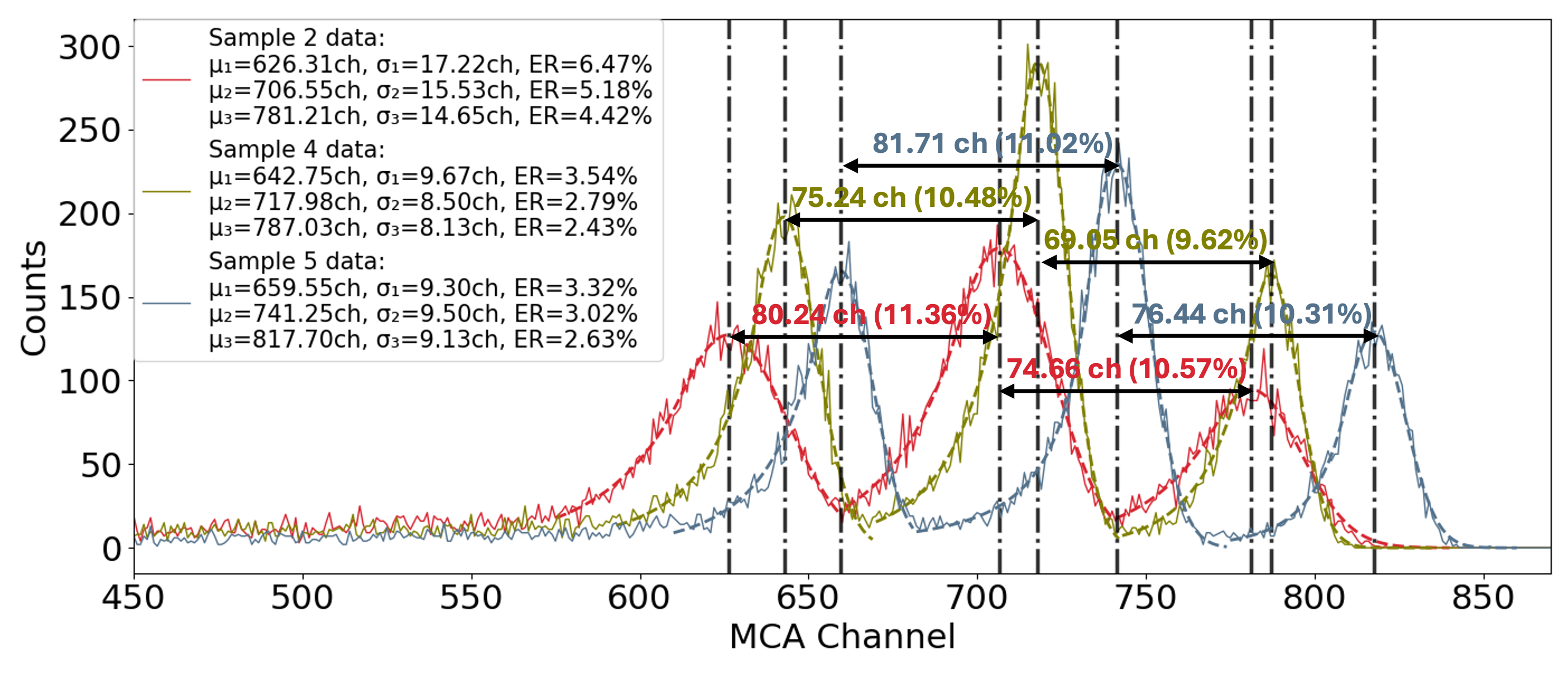}
\caption{Alpha radiation spectra measured with samples 2, 4 and 5 at 650 V bias.\label{fig:Alphas}}
\end{figure}

\begin{figure}[htbp]
\centering
\includegraphics[width=.82\textwidth]{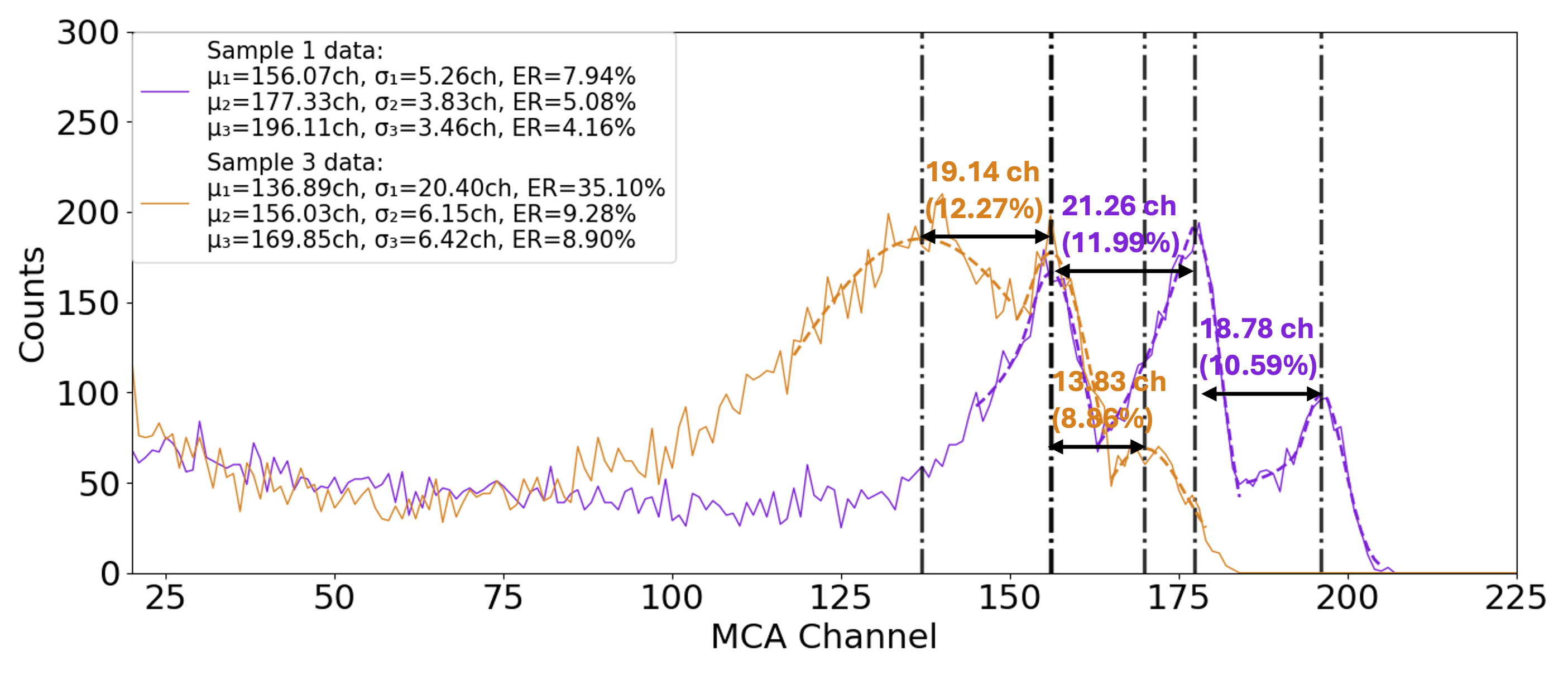}
\caption{Alpha radiation spectra measured with samples 1 and 3 at 650 V bias.\label{fig:Alphas13}}
\end{figure}

Figures \ref{fig:Alphas} and \ref{fig:Alphas13} show the tri-alpha source spectra recorded with the tested samples. The difference in MCA channel values is attributed to the Amptek MCA binning range, which was adjusted due to the observed poor response; otherwise, no changes were made to the setup. The inter-peak distances were calculated relative to the $^{241}\text{Am}$ 5.49 MeV emission centroids.

The three expected emissions, exhibiting the low-energy tailing of planar CZT spectra ~\cite{b}, are observed for all measured samples. Characteristic peak MCA channel position correlates with the sample thicknesses, for thicker samples fixed 650 V bias results in a weaker electric field. However, the alpha radiation response of samples 1 and 3 is substantially worse compared to the rest of the batch. This is reflected in the calculated ER values. Sample 1, compared to sample 3 has more pronounced peaks and visibly deeper valleys between them. Surface or near-surface defects might be a potential reason for deteriorated performance, as the alpha spectral response is highly sensitive to the quality of the surface. Characterization of the impact of defect composition on detection performance is planned to be a subject of future research.

\section{Signal efficiency and \texorpdfstring{$\mu\tau$}{mu-tau} estimation with alpha source}
\label{sec:SE}

To estimate the Signal Efficiency (SE), the recorded waveforms over 2000 events for each bias value were averaged. The waveforms are collected with a trigger of set amplitude, from 2 mV to 4.5 mV. The peak amplitude of the signal waveforms is bias dependent and was recorded to be in the range from 5 mV at 30 V bias to 250 mV at 650V bias. The resulting averaged waveforms for each voltage step are integrated over time. With a sufficiently large data set, the averaged waveform shape remains stable, allowing further data processing. Baseline correction was implemented for all waveforms. 

\begin{figure}[htbp]
\centering
\includegraphics[width=.82\textwidth]{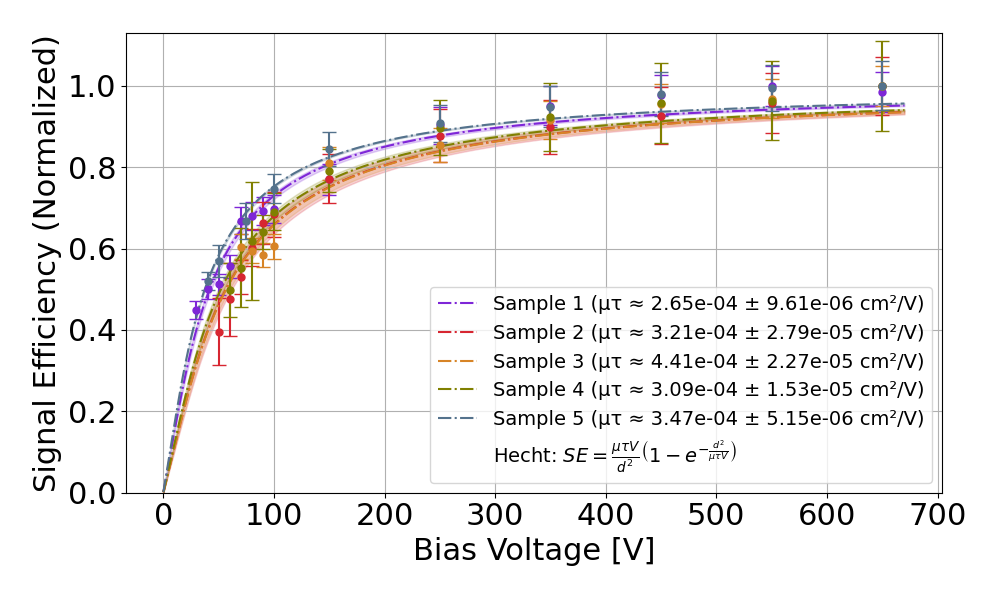}
\caption{Normalized SE - voltage plot for studied samples with the Hecht equation fits used to estimate $\mu\tau$ values. Colored shaded areas on the plot are the uncertainty of the fit. Error bars propagated from the integration limits.\label{fig:SE}}
\end{figure}

A drawback of the used data acquisition method was observed for waveforms recorded at low bias values. Some waveforms tend to be partially negative due to data averaging which directly affects the SE calculation, despite the applied baseline correction. This results in a larger error when calculating $\mu\tau$. The resulting SE plots and calculated $\mu\tau$ values for all characterized samples are shown in figure \ref{fig:SE}.

The signal detected with the CSP is dominated by the electron contribution, as the positive potential was applied to the back electrode, capacitively connected to the signal readout line. Such biasing geometry minimizes the hole-induced signal because positive charge carriers are collected shortly after the alpha particle penetrates the bulk surface levels~\cite{d}. 

A substantial divergence in signal shape formation between the samples is observed with increasing bias voltage. It is valid to assume that the physical properties of sensors, such as resistivity, defect configuration and density alter the response to specific types of ionizing radiation~\cite{e}.

\section{Conclusions}
The described test setups produce reliable spectroscopic and electrical behavior characterizations for CZT measurements. The extracted lifetime-mobility products are consistent with the standard CZT material~\cite{a,d}. This shows that the setup can be used to identify the properties of the sensors, as in this case suggests a common origin for the samples. The bulk resistivity of the samples correlates with the $\mu\tau$ values based on the IV/CV characteristics, except for sample 1 being an outlier. From the obtained values, it is possible to estimate the expected performance of CZT devices. However, in order to have more certainty in the estimation, infrared imaging and follow-up charge-transport studies are required. The defect density and qualitative composition are performance-defining for CZT-based devices~\cite{f}. The developed methods discussed in this paper serve as the basis for subsequent studies linking the detector defect composition to spectroscopic performance.

Large qualitative difference in response to gamma and alpha ionizing radiation for samples 1 and 3 was observed. Based on the IV/CV characteristics and resistivity values, it was expected to see sample 3 perform similarly to the rest of the batch, and the sample 1 to be the outlier. Nonetheless, sample 1 shows better energy resolution under alpha source irradiation, but fails to properly convert the gamma radiation energy to a detectable charge. The described indicates the differences in the surface and bulk defect compositions as a potential reason. These results show that the full set of measurements is required to give a proper estimate of the CZT sample performance.

One further step to improve the SE and $\mu\tau$ estimation results is to record a set of waveforms with sufficient statistics. Filtering out waveforms with specific shapes and amplitudes would allow for a more precise estimation of the SE and provide insight into underlying physical properties, such as defect configuration in the samples.


\begin{thebibliography}{99}

\bibitem{a}
Y. Xiang, X. Jiang, C. Wang, F. Meng, Y. Han, X. Huang and L. Wei,
\emph{Alpha particle detection with a planar CdZnTe detector and relative simulations},
\emph{Radiation Detection Technology and Methods} {\bf 5} (2021) 609–617, doi:10.1007/s41605-021-00296-z.

\bibitem{b}
S.-H. Park, J.-H. Ha, J.-H. Lee, Y.-H. Cho, H.-S. Kim, S.-M. Kang and Y.-K. Kim,
\emph{Fabrication of CZT Planar-Type Detectors and Comparison of their Performance},
\emph{J. Nucl. Sci. Technol.} {\bf 45} (2008) 348–351, doi:10.1080/00223131.2008.10875860.

\bibitem{c}
T.E. Schlesinger, J.E. Toney, H. Yoon, E.Y. Lee, B.A. Brunett, L. Franks and R.B. James,
\emph{Cadmium Zinc Telluride and its use as a Nuclear Radiation Detector Material},
\emph{Mater. Sci. Eng. R} {\bf 32} (2001) 103–189, doi:10.1016/S0927-796X(01)00027-4.

\bibitem{d}
P. Praus, E. Belas, J. Franc, R. Grill, P. Höschl and J. Pekárek,
\emph{Electronic Pulse Shape Formation in Transient Charge and Transient Current Detection Approach in (CdZn)Te Detectors},
\emph{IEEE Trans. Nucl. Sci.} {\bf 61} (2014) 2333–2337, doi:10.1109/TNS.2014.2330070.

\bibitem{e}
A. Hossain, A.E. Bolotnikov, G.S. Camarda, Y. Cui, A. Prishivalko, G. Yang, R. Gul, R.B. James and L. Li,
\emph{Direct observation of influence of secondary-phase defects on CZT detector response},
\emph{J. Cryst. Growth} {\bf 470} (2017) 99–103, doi:10.1016/j.jcrysgro.2016.10.023.

\bibitem{f}
R. Gul, Z. Li, R. Rodriguez, K. Keeter, A. Bolotnikov and R.B. James,
\emph{Defect Measurements in CdZnTe Detectors Using I-DLTS, TCT, I-V, C-V and $\gamma$-ray Spectroscopy},
\emph{Proc. SPIE} {\bf 7079} (2008) 70790U, doi:10.1117/12.797865.

\end{thebibliography}
\end{document}